\documentclass[acmsmall]{acmart}
\usepackage{multirow}
\usepackage{colortbl}
\usepackage{xcolor}
\AtBeginDocument{%
  }

\begin{document}

\title{Back-in-Time Diffusion: \\Unsupervised Detection of Medical Deepfakes}

\author{Fred M. Grabovski}

\email{freddie@post.bgu.ac.il}
\orcid{0009-0000-1652-7614}
\affiliation{%
  \institution{Ben-Gurion University}
  \country{BGU}
}

\author{Lior Yasur}
\affiliation{%
  \institution{Ben-Gurion University}
    \country{BGU}
}
\email{larst@affiliation.org}

\author{Guy Amit}
\affiliation{%
  \institution{Ben-Gurion University}
    \country{BGU}
}

\author{Yisroel Mirsky}
\affiliation{%
  \institution{Ben-Gurion University}
    \country{BGU}
 }
  \authornote{Corresponding Author}

\renewcommand{\shortauthors}{Grabovski et al.}

\begin{abstract}
Recent progress in generative models has made it easier for a wide audience to edit and create image content, raising concerns about the proliferation of deepfakes, especially in healthcare. Despite the availability of numerous techniques for detecting manipulated images captured by conventional cameras, their applicability to medical images is limited. This limitation stems from the distinctive forensic characteristics of medical images, a result of their imaging process.

In this work we propose a novel anomaly detector for medical imagery based on diffusion models. Normally, diffusion models are used to generate images. However, we show how a similar process can be used to detect synthetic content by making a model reverse the diffusion on a suspected image. We evaluate our method on the task of detecting fake tumors injected and removed from CT and MRI scans. Our method significantly outperforms other state of the art unsupervised detectors with an increased AUC of 0.9 from 0.79 for injection and of 0.96 from 0.91 for removal on average. We also explore our hypothesis using AI explainability tools and publish our code and new medical deepfake datasets to encourage further research into this domain.

\end{abstract}




\maketitle
\section{Introduction}
\label{sec:intro}
In recent years, the field of generative AI has gained increasing popularity with the introduction of Generative Adversarial Networks (GANs) and diffusion models which are capable of generating and editing images with impressive quality.
However, these models can be used maliciously to create `deepfakes'; believable  media created by deep neural network \cite{mirsky2021creation}. Deepfakes can be used to spread misinformation~\cite{akhtar2023deepfakes,huh2018fighting}, manipulate forensic evidence~\cite{kim2019deep}, perform blackmail~\cite{Singapor82:online} and even perform scams \cite{frankovits2023discussion}.

The proliferation of deepfakes can be attributed in-part to the growing accessibility of these technologies. Given the tools and online resources today, very little technical expertise is required to create a convincing deepfakes.
For example, with publicly available text-to-image models~\cite{croitoru2023diffusion,ramesh2021zero}, such as Stable Diffusion~\cite{Rombach_2022_CVPR}, anyone can easily create and edit image content using only textual instructions.

Deepfakes can also be applied to the healthcare sector. In a paper called CT-GAN~\cite{mirsky2019ct} the authors demonstrated that malicious actors can gain access to records at a hospital and manipulate CT scans of patients with deepfake content. There are several reasons an adversary would want to perform a deepfake attack on medical imagery.
For example, an attacker can add a small a brain aneurysm to his/her own MRI scan as irrefutable evidence for losing the ability to taste, enabling the attacker to collect money from a quality of life insurance policy. Moreover, an attacker can alter political a candidate's medical records to affect elections or to harm individuals as a form of murder or assassination.

The threat of medical deepfakes is a notable concern because cyber adversaries can access private medical imagery. Numerous research papers and security reports have demonstrated that attackers can access and modify medical scans though cyberattacks and physical intrusions~\cite{beek2018mcafee,eichelberg2020cybersecurity,Abillion19:online,EastRive59:online}. Moreover, advances in generative AI enables adversaries to alter medical evidence in a medical scan with little effort or medical knowledge. For example, it can be used to hide the presence of a cancerous tumor to the degree of realism where both experienced radiologist and AI-tools can be deceived \cite{mirsky2019ct}. Moreover, in this paper we show how text-to-image models such as~\cite{Rombach_2022_CVPR,ruiz2023dreambooth}, can be used to tamper medical imagery by tuning it with just a few images.

A challenge with using existing tamper detection models is that they are designed with the assumption that the images were captured using a conventional camera. Conventional cameras use  CMOS (Complementary Metal-Oxide Semiconductor) or CCD (Charge-Coupled Device) imaging sensors which \textit{linearly} maps received light to pixels in an image. This results in uniform noise patterns in the image which can be utilized to detect image tampering \cite{korus2016multi}.  However, medical imaging devices capture images in a very different manner resulting in non-uniform forensic patterns. For example, CT scanner uses X-rays to capture multiple cross-sectional views of an object from different angles. These views are then processed using a radon transform to reconstruct the internal image (slice). As a result, many existing detection methods, such as \cite{cozzolino2015splicebuster,cozzolino2017prnu,cozzolino2019noiseprint, ghosh2019spliceradar}, cannot detect medical deepfakes very well (demonstrated in section \ref{sec:results}).

Deep learning based anomaly detection is ideal for this setting because unsupervised techniques do not require labeled examples of tampered images; rather examples of benign images which are often plenty. 
Recent advances in generative AI has led to breakthroughs in this domain. Notably, it has been recently shown that Denoising Diffusion Probabilistic Models (DDPMs), which are normally used to generate images, can be used to detect anomalous images as well [][][]. The general approach is to (1) train the DDPM on normal data, (2) noise the target image, (3) use the DDPM to denoise it and then (3) compare the denoised image to the original; if the size of their residual (i.e., their $\ell_2$ distance) is high, then the target image is considered anomalous. 

While the use of DDPMs can help overcome the issue of modeling medical forensics, there are two issues with existing techniques:
\begin{description}
    \item[Content Divergence.] In order to create a residual, existing methods add noise to the image before performing reconstruction. However, as we show in this paper, doing so causes two problems: (1) the DDPM produces novel and irrelevant semantic features which leads to false positives and (2) the magnitude of the added content overpowers subtle forensic patterns discovered in the residual residual. For example, any latent noise patterns from the imaging device and blending artifacts. While this approach works well when searching for anomalous out of distribution content (e.g., an object that was not in the training dataset~\cite{graham2023denoising}) it harms the capability detecting subtle forensic evidence.  Therefore, existing methods are not suitable for deepfake detection and possibly tamper detection in general.
 
    \item[Long Runtime.] When using these techniques, the noised version of the image must be passed through the model multiple times until the noise has been removed. These models are often large and therefore the denoising process of a single image can take tens of seconds and even minutes. While this is not issue when testing a single image, it is a critical one when applying this technique at scale (e.g., when scanning a clinic's entire medical database for tampering).
\end{description}

To address this gap, we introduce a novel DDPM-based anomaly detection method called Back-in-Time Diffusion (BTD). Unlike past works, BTD directly denoises the target image without adding any noise. As a result, the BTD's residual only contains anomalous forensic content. Furthermore, BTD only requires only a single step of diffusion, making it much faster than traditional DDPM processes.

To use BTD, the DDPM model is trained in an unsupervised manner on the target domain, such as random CT scans without tampered content, which eliminates the need for data labeling. This unsupervised approach makes the model practical and cost-effective to train, while also enabling it to generalize well to unseen deepfake technologies, making it robust against a wide variety of manipulation methods.

To thoroughly evaluate our method, we conducted extensive testing across numerous scenarios. We included both MRI and CT medical scans obtained from a diverse range of scanners, ensuring that our model can handle data variability inherent in different imaging devices. We specifically focused on detecting cancer injection and removal manipulations in breast and lung images, as the diagnosis of these conditions are critical. To simulate these deepfakes, we utilized two distinct medical deepfake technologies: CT-GAN and the a generative model based on Stable Diffusion. Finally, for detection we evaluated both conventional image forensic techniques, modern deep learning methods, and DDPM anomaly detectors. 

Our results demonstrate that BTD significantly outperforms all other state-of-the-art detectors across every scenario. For detecting injection deepfakes, BTD achieved an AUC of 0.9, notably higher than the next best method, which only reached 0.79. Similarly, for detecting removal deepfakes, BTD achieved an AUC of 0.96, outperforming the next best method, which scored 0.91. Unlike competing approaches, which show inconsistent performance across different tasks, BTD consistently excels, making it the most robust and reliable detector in all tested cases.

To enable reproducibility of our work and encourage further research,  we published our six novel deepfake datasets and source code online.\footnote{Code: \url{https://github.com/FreddieMG/BTD--Unsupervised-Detection-of-Medical-Deepfakes}\\Datasets: \url{https://www.kaggle.com/datasets/freddiegraboski/btd-mri-and-ct-deepfake-test-sets/}}

\noindent In summary our contributions are:
\begin{itemize}
    \item A novel technique for detecting forensic anomalies in images, evaluated on various medical imaging modalities. The method outperforms other state-of-the-art techniques at detecting medical deepfakes.
    \item We have created and published six new medical deepfake evaluation datasets using state of the art techniques which the community can use as benchmarks.
\end{itemize}

\section{Related Work}
\label{sec:related}
The exploration of tamper detection in images has evolved significantly, initially centering on conventional manipulation techniques like splicing and copy-move~\cite{verdoliva2020media,cozzolino2015splicebuster,lyu2014exposing}. However, the advent of advanced generative models such as GANs and diffusion models has shifted focus towards detecting images synthesized through deep learning~\cite{goodfellow2020generative,creswell2018generative,ho2020denoising}.

Supervised approaches to deepfake detection necessitate extensive training datasets that cover all types of deepfakes~\cite{wu2018busternet,wang2022gan}. This is particularly challenging due to the rapid evolution of deepfake technology and the scarcity of comprehensive datasets, especially in the medical domain.

To counter this challenge, unsupervised learning techniques can be used. Two prevalent methods for detecting image anomalies in an unsupervised setting are: reconstruction-based approaches and fingerprint-based approaches.

\textbf{Reconstruction-based approaches} rely on modeling genuine image distributions and then identifying samples that deviate from them. For example an autoencoder (AE) or Vision Transformer (ViT) can be used to detect manipulated images by identifying cases where the model fails to reconstruct the same image (indicating an outlier)~\cite{zhou2017anomaly,ronneberger2015u,dosovitskiy2020image}. 
Such methods compress and reconstruct the entire image, potentially smoothing over subtle distortions, unlike BTD which effectively captures these distortions by strategically applying only a small number of backward steps in the diffusion process. This approach meticulously peels back the layers of noise added during image manipulation, revealing underlying discrepancies. Using this method, BTD exposes potential tampering, making it adept at identifying manipulations that other methods might miss due to their broader reconstruction approach.

Other methods using diffusion models have also been adapted to the task of anomaly detection. For example, the authors of~\cite{livernoche2023diffusion,graham2023denoising,goodier2023likelihood} utilize these models by performing forward diffusion and backward diffusion, and then by comparing the source image to the reconstructed image. In contrast, BTD only utilizes the backward diffusion for a single step, resulting in faster inference. Furthermore, while these methods perform well at detecting anomalous samples (e.g., novel objects), they perform poorly in the task of detecting image tampering. This is because their backward diffusion process starts from a noisey sample which causes the DDPM to add novel and irrelevant patterns and content to the image. In contrast, BTD extracts the subtle forensic patterns by immediately performing backwards diffusion on the sample, enabling it to detect tampering attacks much better.

Finally, some works suggest using conditional diffusion models for specific anomaly detection tasks~\cite{wolleb2022diffusion,mousakhan2023anomaly}. However, compared to BTD, these approaches are supervised, making them impractical for detecting medical deepfakes. This is because it would require curating and maintaining an up to date large scale medical deepfake dataset.

\textbf{Classic forensic methods} are a common choice for image tamper detection. The popular approach is to search the media for the latent noise pattern left by the imaging device. Tampering is detected if the latent noise pattern has been disturbed due to blending (splicing) or removed due to synthesized content~\cite{chen2008determining, mareen2022comprint, cozzolino2017prnu}.
These approaches assume that images contain uniform noise patterns left by the imaging sensor~\cite{chen2008determining, ghosh2019spliceradar, cozzolino2019noiseprint}.
However, the imaging processes of medical devices (such as MRI and CT) is different than in standard cameras for which these algorithms were designed. For example, CT and MRI scanners have an imaging head that spins around the object while performing radial integrals to visualize the object volume; a process which results in \textit{non-uniform} latent noise patterns.

Other works offer methods for detecting tampering by comparing an image to its metadata such as exposure and ISO settings~\cite{huh2018fighting}. Unfortunately, these approaches cannot be directly applied to medical imagery.
In contrast, our Back-in-Time Diffusion (BTD) technique leverages the unsupervised potential of Denoising Diffusion Probabilistic Models (DDPMs) to offer a versatile solution for detecting manipulated images, bypassing the limitations of existing methods in the medical domain and in additional scenarios in which the imaging process is different or the availability of image meta data is limited.

\section{Back-In-Time Diffusion}
\label{sec:diffusion_models}

\subsection{Diffusion Models}
Denoising diffusion probabilistic models (DDPM)~\cite{ho2020denoising} are a class of generative models which are inspired by concepts from nonequilibrium thermodynamics.

DDPMs utilize two processes (1) forward diffusion, which is the process of noising an input sample $x_0 \sim q(x_o)$ into a noise $x_T \sim \mathcal{N}(0, I)$, and (2) the backward diffusion process, that reverses the forward diffusion process, from $x_T$ to $x_0$ using $T$ time steps.

The training of DDPMs is an iterative process, where in each iteration, a random number of steps $t$ is selected, and the forward diffusion process is used to noise an input sample $x_0$ into $x_t$ over $t$ steps using a closed mathematical formula
\begin{equation}
    x_{t} = \sqrt{\Bar{\alpha_t}}\cdot x_0 + \sqrt{1-\bar{\alpha_t}} \cdot \epsilon_t
\end{equation}
where $\epsilon_t \sim \mathcal{N}(0, I)$ and $\bar{\alpha_t} = \prod_{i=1}^t\alpha_i$ is a term responsible for controlling the image content loss in each time step of the forward process.
After the noised sample $x_t$ is calculated, the backward diffusion process uses a U-Net-like~\cite{siddique2021u} neural network $F(x_t)$ to estimate the noise $\epsilon_t$ between $x_t$ and $x_{t-1}$.
The neural network is then updated according to a gradient step minimizing the squared difference between the noise $\epsilon_t$ and the neural network output $F(x_t)$:
\begin{equation}
    loss = \lvert| \epsilon_t - F(x_t) |\rvert^2
\end{equation}

Finally, after the neural network's training has converged, the backward diffusion process can be used to generate new images.
The process starts from noise $\epsilon\sim \mathcal{N}(0, I)$ and then repeatedly estimates $x_{t-1}$ from $x_t$, until $t=0$, resulting in a sample $x_0$ similar to samples in the training set.
More details about diffusion models can be found at~\cite{ho2020denoising,croitoru2023diffusion}.

\subsection{The Problem with Existing DDPM Anomaly Detection}
\label{sec:approach}

\begin{figure*}[bt]
    \centering
    \includegraphics[width=\columnwidth]{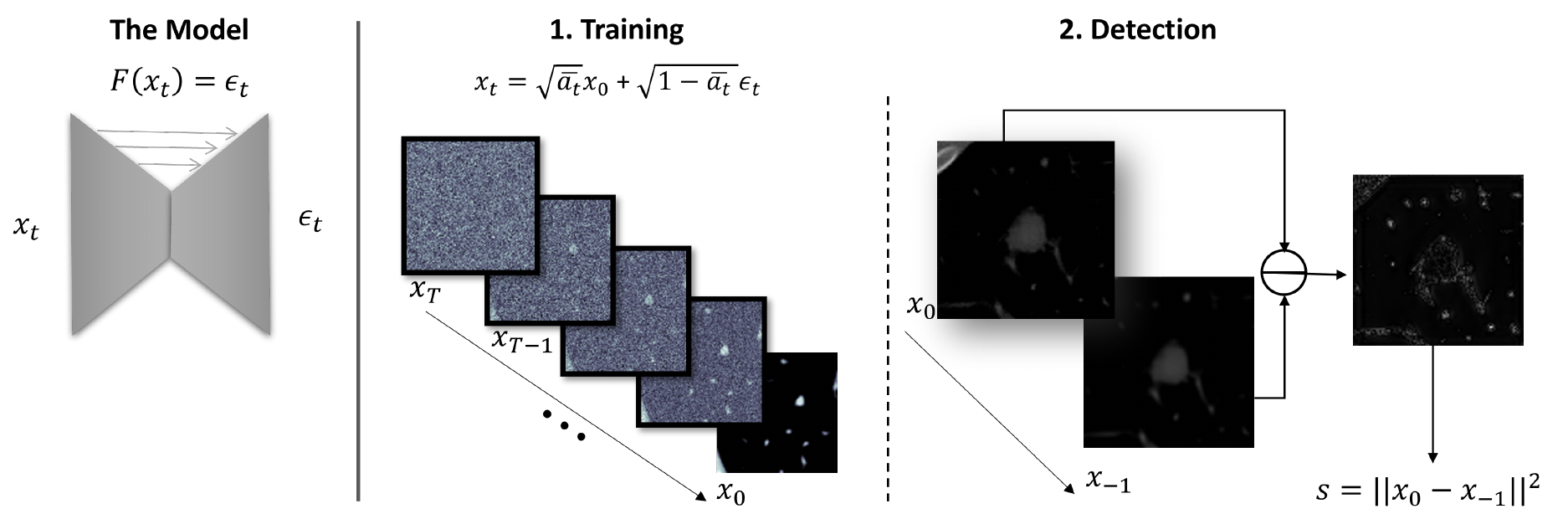}
    \vspace{-1em}
    \caption{Overview of the Back-in-Time Diffusion framework.
    The model is a U-net that predicts the noise added in the last step ($F(x_t) \Rightarrow \epsilon_t$).
    The model is trained by progressively adding noise to the images \((x_0)\) to generate noisy versions \((x_t)\) and then learning to predict the noise added at each step.
    Detection is performed on image $x_0$ by measuring the error after $d$ steps.}

    \label{fig:BTD}

\end{figure*}

Let $x_0$ be a target image which we want to examine for anomalies. 
Existing DDPM anomaly detectors work by using the DDPM process to remove anomalous features and then by comparing the result. The process is to first train a DDPM \textit{on authentic images only}. Then the DDPM is used to determine if $x_0$ is anonymous by following three steps: 
\begin{enumerate}
    \item \textit{Forwards diffusion.} $x_0$ is noised $k$ steps to remove the anomalous content.
    \item \textit{Backwards diffusion.} $x_k$ is denoised $k$ steps to produce $x'_0$; the reconstruction of $x_0$ without anomalous content.
    \item \textit{Residual.} The magnitude of the residual $||x_0 - x'_0||^2$ is computed to quantify the anomalous content in $x_0$.
\end{enumerate}

The intuition is that since the backwards diffusion model was only trained on benign content, it will only be able to repair $x_k$ with normal features. As a result, the residual will capture what is anomalous w.r.t. the training distribution.

The problem with this process is that the forwards diffusion steps add add to the image. This encourages the backward diffusion process to create novel and irrelevant content to the image. This leads to increased false positives and the inability to measure the presence of subtle forensics because the magnitude of these novel features.

\subsection{The Proposed Approach (BTD)}
In order to avoid the issue of generating irrelevant content, we instead opt for utilizing DDPMs to identify subtle forensics left in the image from the tampering process. Examples of these forensics include, blending boundaries and inconsistent latent noise patterns~\cite{verdoliva2020media,piva2013overview, su2024ufcc}. 

To accomplish this, we do not add any noise to $x_0$ but rather perform a single denoising step directly to it. By doing so, the model attempts to correct the latent noise patterns to make them appear more similar to samples from the training set. As a result, the residual captures \textit{anomalous} noise patterns apparent in the image. Therefore, the BTD process has two steps in total (illustrated in Figure~\ref{fig:BTD}):

\begin{enumerate}
    \item \textit{Backwards diffusion.} Target image $x_0$ is denoised for one step obtaining $x_{-1}$. 
    \item \textit{Residual.} The magnitude of the residual $||x_0 - x_{-1}||^2$ is computed to quantify the anomalous content in $x_0$. A larger residual indicates a higher likelihood that the image contains anomalous noise patterns.
\end{enumerate}

While it is possible to continue the backward diffusion process further than one step, we found that doing so harms performance. This is because doing so harms both the benign subtle features and eventually benign foreground features; leading to false positives.

Since by definition $F(x_0) \approx x_{-1} -x_0$, the entire process can be summarized as
\begin{equation}
    s = ||F(x_0)||^2
\end{equation}
where $s\in [0,\infty)$ is the anomaly score.

For medical deepfakes, we found that when content (i.e., tumors) are injected into a scan, it is beneficial to focus on the residual around the tampered area itself. Therefore, when looking for injected content, we take the average around the area of interest. Specifically, we take the average of the residual 32x32 patch around the center of the tumor.

\subsection{Calibrating BTD}
The identification of fake samples is performed by flagging samples that obtain an anomaly score greater than some predetermined threshold $\tau$:
\begin{equation}
    p(score) = \begin{cases}
     \text{Real} \;\; \text{if} \quad \tau\leq s
        \\
        \text{Fake} \;\; \text{if} \quad \tau > s
        \end{cases}
\end{equation}
The threshold $\tau$ can be determined statistically in an unsupervised manner: first we compute the anomaly scores of benign images that were nit used to train $F$. Then fit the distribution to a CDF and set the $\tau$ such that the probability of a false positive it sufficiently low.

\subsection{Detection Stability} 
DDPMs are known for their stochastic nature, attributed to the random noise sampled during the backward diffusion process. Deploying a DDPM based detection tool may raise some understandable concerns regarding the stability and reproducibility of detection performance. However, in our experiments we have found that since we only take one backwards step, there is a negligible variance in reconstruction results. Therefore, BTD can offer reliable results.

\section{Experiment Setup}
In this section, we describe how we evaluated BTD's ability to detect deepfakes. For reproducability, we have made the the source code of BTD and the datasets used in this paper available online.\footnote{\url{https://github.com/FreddieMG/BTD--Unsupervised-Detection-of-Medical-Deepfakes}\\\url{https://www.kaggle.com/datasets/freddiegraboski/btd-mri-and-ct-deepfake-test-sets}}

\noindent{Attack Scenarios.} We considered different medical modalities, attacks scenarios and medical deepfake technologies:

For the modalities, we examined cancerous tumors in CT lung and MRI breast scans. For the attack scenarios, we analyzed the injection and removal of tumors. In the case of injection, samples representing authentic images with malignant tumors are called True-Malign (TM) samples, and those with artificially inserted malignant tumors are called Fake-Malign (FM) samples. Similarly, for removal, True-Benign (TB) samples are genuine benign images and Fake-Benign (FB) samples are images with artificially removed tumors. For the medical deepfake technologies, we used CT-GAN from~\cite{mirsky2019ct} which performs volumetric in-painting and Stable Diffusion (SD) from~\cite{Rombach_2022_CVPR,ruiz2023dreambooth} to tamper the images. We implemented the SD injection attack by fine-tuning a Stable Diffusion model specifically on malignant slices (one for each modality). For the removal of tumors using Stable Diffusion, we utilized an "off-the-shelf" model without the need for additional fine-tuning.
To the best of our knowledge, this is the first work that demonstrates the use of SD for creating medical deepfakes. 

\subsection{Datasets}
To create our datasets we first collected authentic TM and TB images from the Duke Breast Cancer MRI dataset~\cite{saha2021dynamic} and the LIDC ~\cite{armato2015data} datasets. MRI images were standardized on a per-patient basis and CT images were clipped between -700 (Air) and 2000 (Bone).
Both image modalities where then scaled to a [0, 1] range. Next, we preprocessed the datasets by cropping the imagery around regions of interest. For the TM dataset, this involved focusing on annotated tumor areas, while for the TB dataset, healthy tissue regions were randomly selected from the list of tumor locations to reflect the potential for containing a tumor. We then split these datasets into two portions: one for training  BTD and a hold-out set both used for validating BTD and tuning the medical deepfake models (Stable Diffusion).

\begin{figure*}[h]
    \centering
    \includegraphics[width=0.65\textwidth]{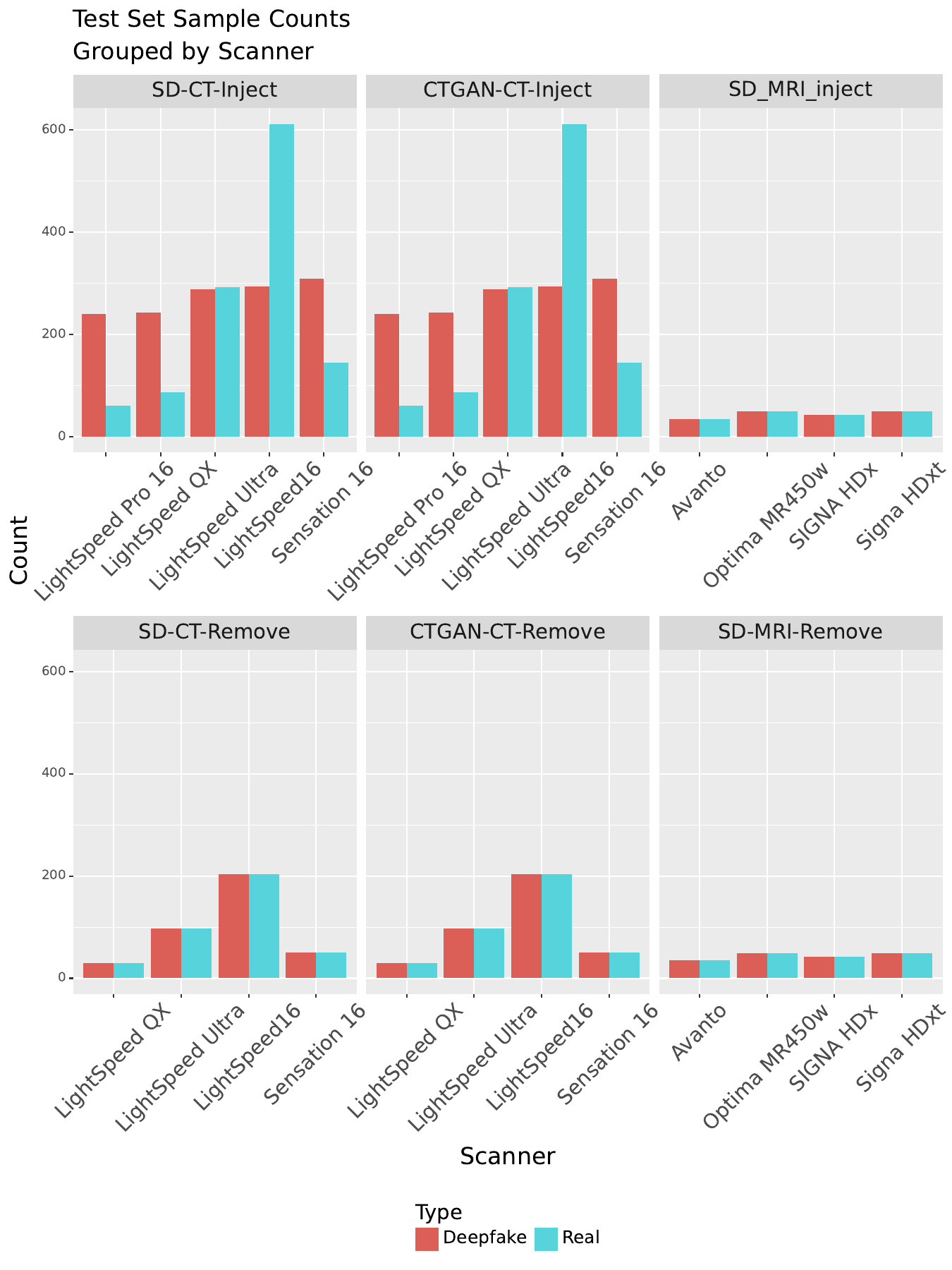}
        \vspace{-1em}
    \caption{The number of samples from each imaging device, per dataset.}
    \label{fig:test_sets}

\end{figure*}

\begin{figure*}[bt]
\centering
    \includegraphics[width = \textwidth]{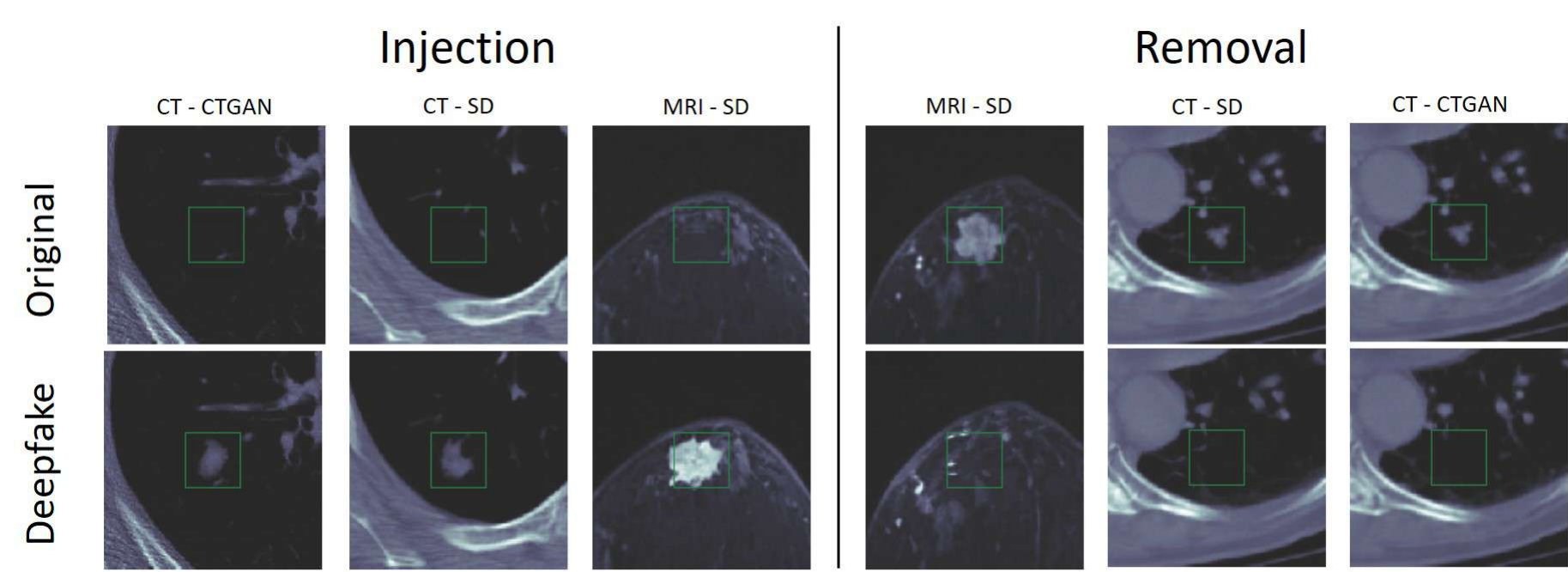}
\caption{Examples of manipulated images from our detection datasets. Top: original images. Bottom: images after being edited by generative models}
\label{fig:AI_tampering_showcase}

\end{figure*}

Utilizing the images in the hold-out set, we compiled six datasets for the evaluation: \texttt{CTGAN-CT-Inject}, \texttt{CTGAN-CT-Remove}, \texttt{SD-CT-Inject}, \texttt{SD-CT-Remove},  \texttt{SD-MRI-Inject}, and \texttt{SD-MRI-Remove}. These datasets consist of images from five different CT scanners and four MRI scanners. 
Sample images from our datasets are provided in Figure~\ref{fig:AI_tampering_showcase} where we show the before and after of applying the attacks. In Figure~\ref{fig:test_sets} we provide distribution of these data sets.

In total, we use 77K CT
slices and 47K MRI slices for training, and for validation,
we use 17K CT slices and 14K MRI slices. To avoid bias, the split was done according to patient. The train and validation sets only contained true samples (TB, TM) while the test sets contained a mix (TB, TM, FM and FB). The detectors were then trained using the train and validation sets and then evaluated on the respective test sets.

\subsection{BTD Configuration}
Our framework revolves around a UNET-based Diffusion model, for which we used the default hyper-parameters,~\footnote{https://github.com/lucidrains/denoising-diffusion-pytorch} with the exception of the model's dimensions and batch size, which we changed to 32 \texttt{init\_features} and 64 respectively.
To apply BTD, we first extract a patch around the area interest and then pass it through our model. We set the patch sizes to 96x96 and 128x128 for CT and MRI respectively to encapsulate potential tumors with sufficient context.



\subsection{Baselines}
Not all detection methods are applicable to our setup.
To ensure a fair comparison, we only used methods that (1) train their model in an unsupervised manner, and (2) do not require additional metadata that is typically unavailable in medical imaging datasets.
We compare BDT to the following methods:

\begin{description}
    \item[Autoencoder Variants.] AEs are frequently used for image anomaly detection tasks~\cite{yang2021visual}.
    In general, when using AEs for image anomaly detection tasks, the original image is compared to its reconstructed version resulting in a map which localizes anomalous regions.
    We adopted four variants of AEs in our evaluation:
    \begin{itemize}
        \setlength{\itemsep}{0pt}%
        \setlength{\parskip}{0pt}%
        \item \texttt{AE}~\cite{zhou2017anomaly} - A standard convolutional AE with 4 encoder and 4 decoder layers.
        \item \texttt{U-net}~\cite{ronneberger2015u} - An AE based on the U-net architecture, with 16 \textit{init\_features}.  
        \item \texttt{In-paint AE}~\cite{haselmann2018anomaly} - An in-painting AE for anomaly detection. During inference, the center part of the image is masked and the AE attempts to complete it.
    
        \item \texttt{OC-VAE}~\cite{khalid2020oc} - A Variational AE architecture that was originally proposed for detecting fake face images.
    \end{itemize}

    \item[Splice Buster.]~\cite{cozzolino2015splicebuster} A common baseline in the field of image forensics. 
    This technique segments images into smaller sections and isolates the inherent noise present in each segment during the imaging procedure.
    These segments are subsequently grouped based on their distinctive noise patterns using a Gaussian Mixture Model, leading to two categories: benign and fake.
    The assigned probabilities for each segment are then employed to construct a heat map that indicates the likelihood of each pixel within the image being counterfeit.
    
    \item[Splice Radar.]~\cite{ghosh2019spliceradar}: An advanced method in image forensics, utilizing a deep neural network (DNN). 
    The approach involves training the DNN to grasp image characteristics, specifically the image fingerprint.
    This is accomplished by instructing the DNN to differentiate between the source imaging devices of the training images. 
    Notably, the initial layer of the DNN is restricted to capturing solely high-frequency attributes from the image input. 
    After training the DNN, it extracts representations for individual image segments.
    These representations are then inputted into a Gaussian Mixture Model, which clusters the segments into groups similar to SpliceBuster's technique.  
    
    \item[Satellite ViT.]~\cite{dosovitskiy2020image,horvath2021manipulation}: Designed to detect image manipulations using Vision Transformers (ViT)~\cite{dosovitskiy2020image}.
    The fundamental concept driving this approach is rooted in the distinction of derivative patterns between benign and fake images.
    To this end, the method employs a ViT which is used as an AE.
    For every given input image, a comparison is performed between its derivative and the ViT's output corresponding derivative, facilitating the generation of a heat map that indicates the specific areas containing fake content within the image. We calculate thresholds from the validation sets of CT and MRI to keep only the more 'active' sections of the heat map and sum to calculate an anomaly score

\item[OOD DDPM.]~\cite{graham2023denoising}: An Out-of-distribution (OOD) detection method based on DDPMs. It detects anomalies by (1) reconstructing the target image from varying noise levels and then (2) by measuring the residuals of the reconstructions. Higher errors suggest the input differs from the training data, indicating an anomaly. In contrast to BTD, this detector must add noise to the images to perform the detection process.

\end{description}

All the methods in the evaluation were applied to our data datasets.
We have used the author's code wherever possible, and in cases where the code was not available, we managed to get in touch with the authors in order to reproduce their work correctly.


Two models were trained for each method, one for CT and one for MRI. Each model was trained using benign and malignant image slices without tampered content.

\subsection{Metrics} Performance was measured using area under the curve (AUC) and equal error rate (EER). For AUC, a larger value is better where 1.0 and 0.5 indicate perfect classification and random guessing respectively. For EER, a lower value indicates better performance. Our evaluation metrics are calculated using bootstrap sampling with 100 iterations, where in each iteration the number of samples per class were balanced across the scanners in the dataset. The results presented in the paper are the means calculated across the 100 iterations.

\section{Results}
\label{sec:results}

In this section, we evaluate the performance of BTD in detecting deepfakes in medical images.
We start by comparing BTD to the baselines and analyze how well the method generalizes across different imaging machines (makes and manufacturers). Finally, we provide an ablation study on BTD to (1) confirm our hypothesis that forwards diffusion harms detection performance and (2) confirm that one step of backwards diffusion (i.e., $k=1$) is ideal.

\subsection{Baseline Analysis}

In Table~\ref{tab:main_results} we present the detection performance of two BTD detectors using different parameters (\texttt{init\_features} of 8 and 32). The table shows that BTD outperforms the other baselines with a significant margin. This is regardless of whether the baseline was designed to detect semantic anomalies (e.g., the AE variants) or classical image forensic anomalies (Splice detectors and Satellite ViT). 
The exception to this observation is on the dataset (\texttt{CTGAN-CT-Remove}) where \textbf{all} of the methods fail, including BTD. For this dataset, every method obtains an AUC of about 0.55 which is essentially random guessing.
We believe that the reason for this is that there are very few semantic anomalies left behind after a removal attack. Therefore, the models must rely on forensic anomalies alone to identify tampering (e.g., noise patterns). However, in this dataset, CT-GAN uses noise to further mask the removal of tumors. This is why BTD also struggles to identify the attack.

\begin{figure}[t]
    \centering
    \includegraphics[width=\textwidth]{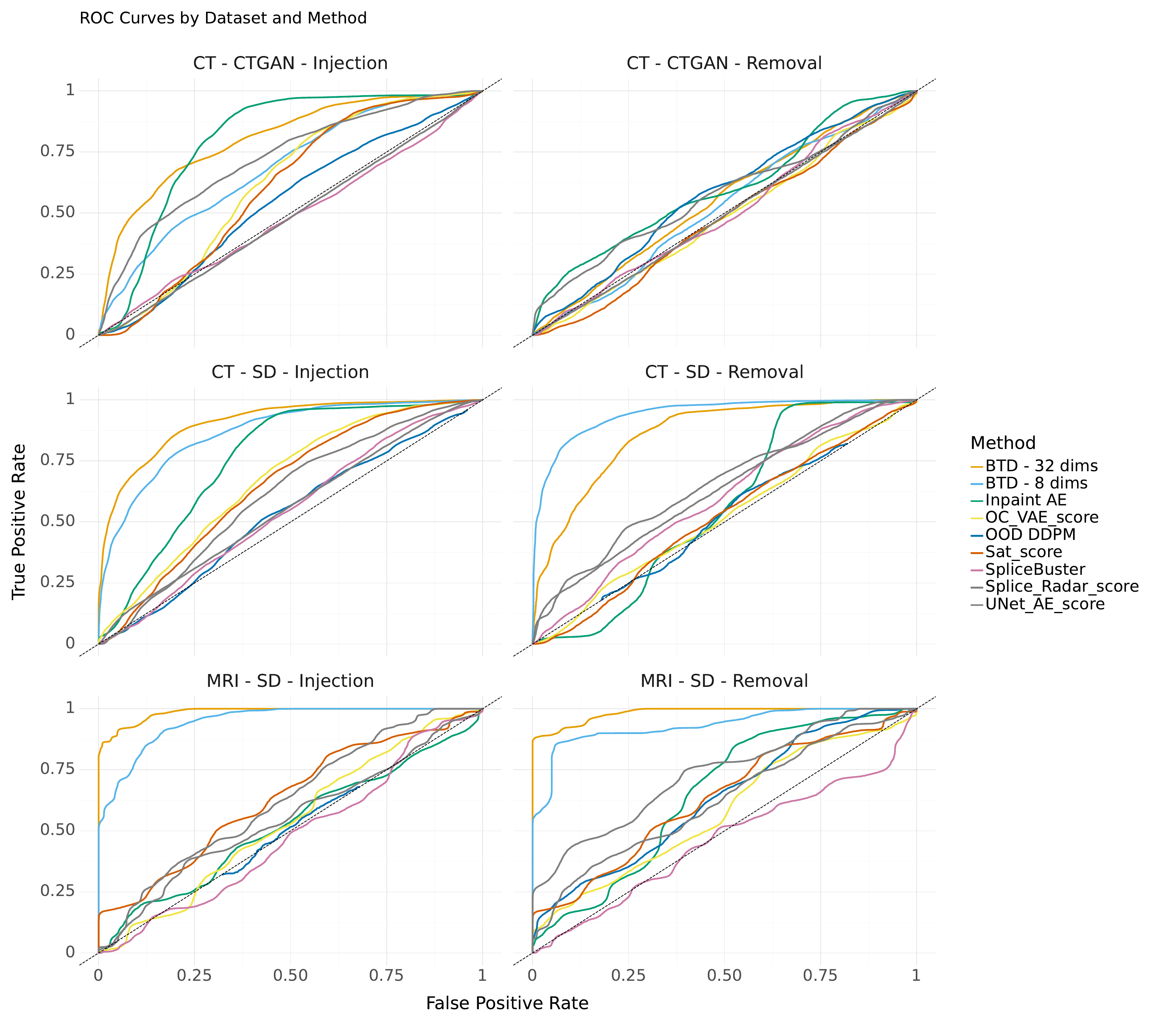}
    \caption{ROC Curves of BTD and the baselines organized by dataset (attack scenario).}
    \label{fig:roc_curves}
\end{figure}

False positive rates are crucial in anomaly detection, particularly in the medical domain. To better understand the sensitivity of these algorithms, we plot their receiver operating characteristic (ROC) curves in Figure~\ref{fig:roc_curves}. The plot demonstrates that by adjusting the threshold of BTD, it is possible to achieve higher true positive rates with a very low false positive rate compared to other methods. Additionally, when stable diffusion is used to perform the attack, BTD can achieve a zero false positive rate.

\begin{table}[t]
\caption{Baselines comparison: The AUC of BTD compared to the baselines for each of the six attack scenarios. Bold values indicate the best result over 0.6 (close to random guessing).}
\label{tab:main_results}
\begin{tabular}{c|ccc|ccc|}
\textit{\textbf{}} & \multicolumn{3}{c|}{Injection} & \multicolumn{3}{c|}{Removal} \\ \cline{2-7} 
 & MRI & \multicolumn{2}{c|}{CT} & MRI & \multicolumn{2}{c|}{CT} \\ \cline{2-7} 
 & SD & CT-GAN & SD & SD & CT-GAN & SD \\ \hline
AE & 0.80 & 0.56 & 0.79 & 0.91 & 0.45 & 0.93 \\
In-Paint AE & 0.52 & 0.80 & 0.76 & 0.65 & 0.60 & 0.56 \\
OC-VAE & 0.54 & 0.62 & 0.68 & 0.58 & 0.48 & 0.52 \\
Unet-AE & 0.63 & 0.46 & 0.71 & 0.73 & 0.48 & 0.92 \\
Splice Buster & 0.48 & 0.49 & 0.54 & 0.46 & 0.50 & 0.59 \\
Splice Radar & 0.55 & 0.73 & 0.61 & 0.62 & 0.57 & 0.65 \\
Satellite ViT & 0.54 & 0.61 & 0.65 & 0.63 & 0.47 & 0.51 \\
OOD DDPM & 0.50 & 0.54 & 0.53 & 0.63 & 0.42 & 0.49 \\
\rowcolor[HTML]{EFEFEF} 
\textbf{BTD} - 32 dim & \textbf{0.99} & \textbf{0.81} & \textbf{0.91} & \textbf{0.98} & 0.55 & 0.86 \\
\rowcolor[HTML]{EFEFEF} 
\textbf{BTD} - 8 dim & 0.93 & 0.69 & 0.87 & 0.93 & 0.52 & \textbf{0.94} \\ \hline
\end{tabular}%
\end{table}


\subsection{Generalization}

The medical images can be obtained using a wide variety of scanners from different manufactures. Each device may leave a different forensic signature or even capture the media with various levels of quality. Therefore, it is important to analyze the generalizability of BTD across different settings and scenarios, as opposed to looking at its average performance as a whole.

In our experiments, we used four different types of LightSpeed  scanners and one type of Sensation scanner for the CT modality. For the MRI modality we used one type of scanner for Avanto and Optima, and two type of scanners for SIGNA. In Table \ref{Table:Scanners} we provide the AUC and EER of each scanner for each dataset and attack. Our findings indicate that BTD is a robust detection method regardless of the imaging scanner or technology used. The only exception is in the case of the removal attack on CT scans, as previously explained.

\begin{table*}[ht]
\centering

\caption{The performance of BTD when applied to images taken from different medical imaging devices.}
\label{Table:Scanners}
\resizebox{\textwidth}{!}{
\begin{tabular}{ccc}
    \begin{tabular}[t]{ccc}
        \multicolumn{3}{c}{SD-CT-Inject}\\
        \hline\hline
        \textit{Scanner} & \textit{AUC} & \textit{EER} \\ \hline
        LightSpeed16 & 0.93 & 0.15 \\
        Sensation 16 & 0.94 & 0.10 \\
        LightSpeed Pro 16 & 0.93 & 0.18 \\
        LightSpeed Ultra & 0.85 & 0.22 \\
        LightSpeed QX & 0.89 & 0.17 \\
        \hline\hline
    \end{tabular}
    &
    \begin{tabular}[t]{ccc}
        \multicolumn{3}{c}{CTGAN-CT-Inject}\\
        \hline\hline
        \textit{Scanner} & \textit{AUC} & \textit{EER} \\ \hline
        LightSpeed16 & 0.75 & 0.28 \\
        Sensation 16 & 0.96 & 0.10 \\
        LightSpeed Pro 16 & 0.90 & 0.18 \\
        LightSpeed Ultra & 0.66 & 0.42 \\
        LightSpeed QX & 0.69 & 0.38 \\
        \hline\hline
    \end{tabular}
    &
    \begin{tabular}[t]{ccc}
        \multicolumn{3}{c}{SD-MRI-Inject}\\
        \hline\hline
        \textit{Scanner} & \textit{AUC} & \textit{EER} \\ \hline
        Avanto & 0.92 & 0.20 \\ 
        Optima MR450w & 1.00 & 0.00 \\ 
        SIGNA HDx & 1.00 & 0.00 \\ 
        Signa HDxt & 1.00 & 0.01 \\ 
        \hline\hline
    \end{tabular}
    \\
    \\

    \begin{tabular}[t]{ccc}
        \multicolumn{3}{c}{SD-CT-Remove}\\
        \hline\hline
        \textit{Scanner} & \textit{AUC} & \textit{EER} \\ \hline
        LightSpeed16 & 0.92 & 0.15 \\ 
        Sensation 16 & 0.93 & 0.14 \\ 
        LightSpeed Ultra & 0.85 & 0.23 \\ 
        LightSpeed QX & 0.78 & 0.28 \\
        \hline\hline
    \end{tabular}
    &
    \begin{tabular}[t]{ccc}
        \multicolumn{3}{c}{CTGAN-CT-Remove}\\
        \hline\hline
        \textit{Scanner} & \textit{AUC} & \textit{EER} \\ \hline
        LightSpeed16 & 0.56 & 0.45 \\
        Sensation 16 & 0.56 & 0.44 \\
        LightSpeed Ultra & 0.58 & 0.44 \\
        LightSpeed QX & 0.52 & 0.48 \\
        \hline\hline
    \end{tabular}
    &
        \begin{tabular}[t]{ccc}
        \multicolumn{3}{c}{SD-MRI-Remove}\\
        \hline\hline
        \textit{Scanner} & \textit{AUC} & \textit{EER} \\ \hline
        Avanto & 0.97 & 0.09 \\ 
        Optima MR450w & 1.00 & 0.02 \\ 
        SIGNA HDx & 1.00 & 0.00 \\ 
        Signa HDxt & 1.00 & 0.00 \\ 
        \hline\hline
    \end{tabular}
\end{tabular}
}

\vspace{-1em}
\end{table*}

\subsection{Ablation Study}

In this section, we analyze the contribution of BTD's direct backward diffusion step and other aspects which affect BTD's performance.

\subsubsection{Eliminating Forwards Diffusion}
In section \ref{sec:approach} we hypothesized that other diffusion based anomaly detectors fail because they perform forwards diffusion during the detection step. As a result, these methods generate irrelevant content during the backwards diffusion steps which leads to false positives and overpowers signatures of any subtle forensic evidence. However, BTD does not have this issue because it only performs backwards diffusion.

To validate this claim, we compare both approaches with increasing numbers of diffusion steps. We try in the backwards direction and we also try in both directions. Table \ref{tab:noise_ablation} shows that in every case, applying the forwards diffusion process harms the results, even with a few steps. Therefore, performing only backward diffusion (BTD) is preferred. As discussed earlier, While the combine approach works well at identifying novel object and semantics, a fake cancer still appears as cancer. Thus, by performing backwards diffusion only, we have a better chance at detecting the tampering because we focus on the subtle forensics only such as distortions around the blending boundaries or latent noise patterns around the cancer.

To support this observation, we performed an empirical experiment to assess why the BTD residuals are more informative. We use an explainable AI tool called SHAP \cite{lundberg2017unified} to understand what anomaly whould look at in both cases. SHAP's DeepExplainer provided insights into the contributions of individual pixels (or regions) to the model's predictions. This allows us to visualize and quantify the significance of areas in the residual that contribute tot he model's decision.

To conduct this experiment, we performed the following: from the MRI datasets \texttt{SD-MRI-Inject} and \texttt{SD-MRI-Remove}, we derived two different sets of residuals. The first was generated using the BTD approach ($|x_0 - x_{-1}|$), which involves only a single backward diffusion step. The second version performed 20 forward and then 20 backward steps, producing residuals ($|x_0 - x'_{-20}|$) for comparison.
For each approach, we trained a convolutional neural network (CNN) to distinguish between real and fake residuals. These residuals were derived from images that the diffusion process had not encountered during training. We then ran the SHAP DeepExplainer tool on these residuals and analyzed the results.

Overall, we found that the SHAP values from the forwards-backwards approach yielded higher entropy compared to the BTD method. Specifically, the entropy of SHAP values for BTD residuals was 2.18 in the removal scenario and 2.20 in the injection scenario, whereas the entropy for forwards-backwards residuals averaged 2.94 for removal and 2.75 for injection. This indicates that the BTD method provides more informative insights overall, as lower entropy suggests more focused and interpretable contributions in the model's decision-making process.

\begin{figure*}[t]
  \centering
    \includegraphics[width=1\textwidth]{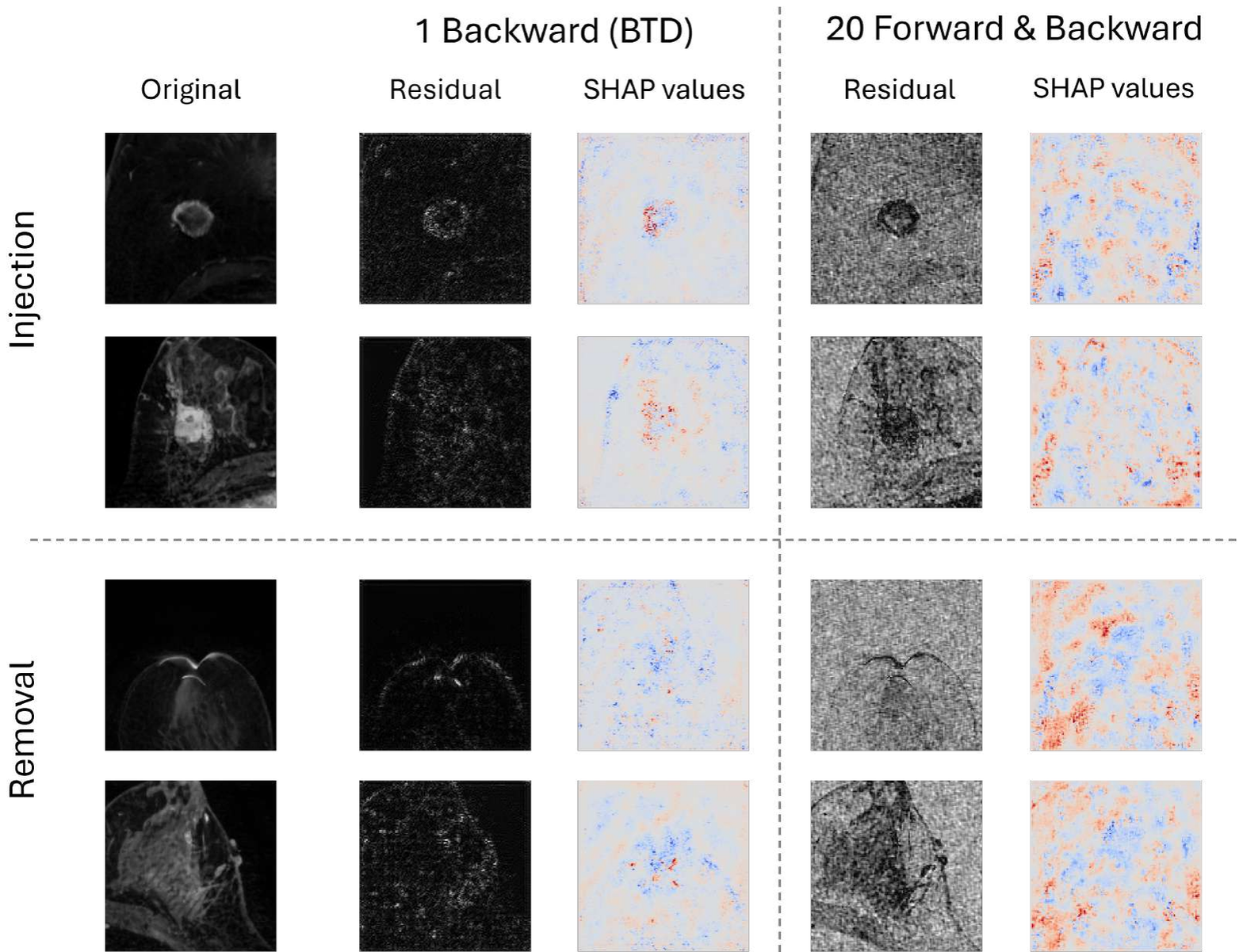}

  \caption{BTD Explainabiliy Experiment: BTD residuals and their SHAP values compared to residuals created with 20 F\&B and their SHAP values}
  \label{fig:SHAP}
\end{figure*}

A visual representation from this experiment supports our findings. Figure~\ref{fig:SHAP} demonstrates that the residuals from the BTD approach provide more concentrated information around the tampered areas (e.g., the center of the image). This focused interpretation aligns with BTD's objective of isolating forensic evidence indicative of tampering, enhancing its effectiveness in detecting manipulations. In contrast, the SHAP values from the forwards-backwards approach are more dispersed and less focused. This scattering suggests that the model struggles to identify relevant forensic features and instead relies on irrelevant noise and generated content. This unfocused behavior supports our hypothesis that the forward diffusion steps introduce redundant content, leading to less reliable anomaly detection and obscuring subtle evidence of tampering.

\subsubsection{The Number of Backward Steps}
\begin{figure*}[t]
  \centering
    \includegraphics[width=0.49\textwidth]{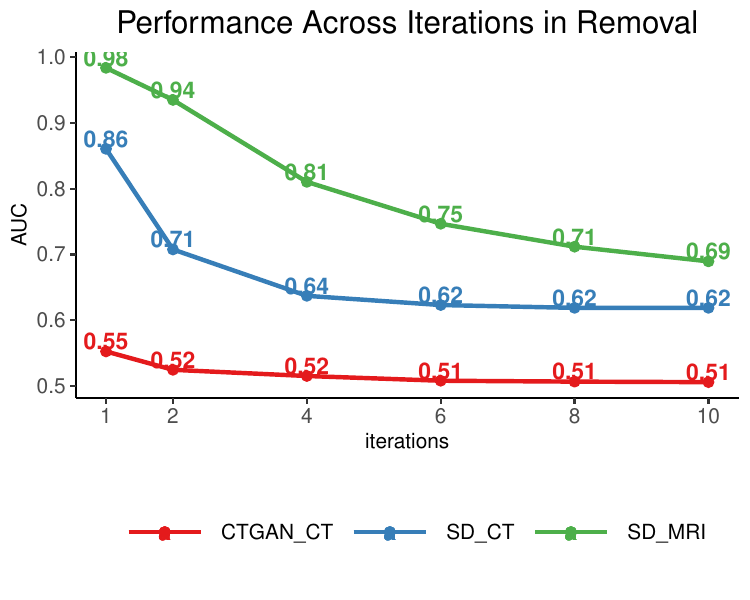}
    \includegraphics[width=0.49\textwidth]{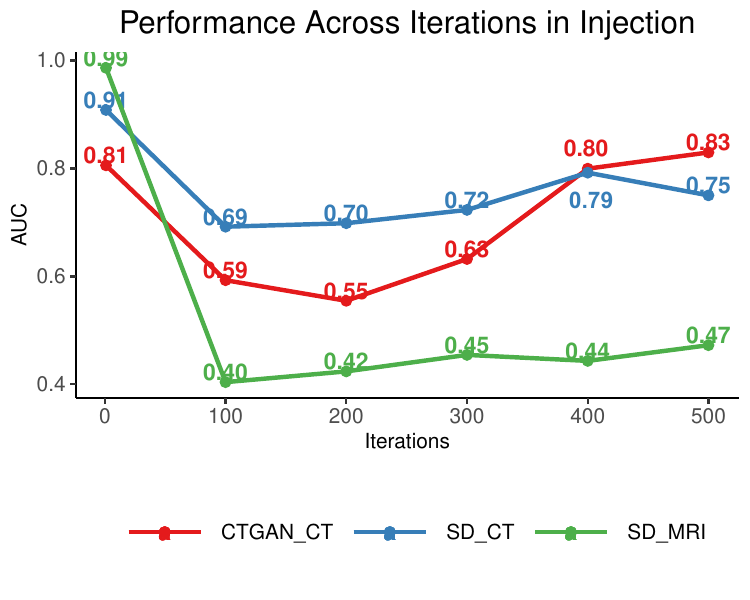}
  \caption{The impact of diffusion step count on BTD performance for removal (left) and injection (right).}
  \label{fig:t_analysis}
\end{figure*}

The hyper-parameter $d$ is the number of 
backward diffusion steps performed by BTD. In Figure~\ref{fig:t_analysis} we plot its impact on detection performance for each of the datasets. We found that more than one backwards diffusion step harms detection performance of both removal am injection attacks. In the case of detecting injections, further steps raise performance slightly but platues at $d=400$ and does not surpass $d=1$. We believe this pattern is because there are some larger forensic patterns that take several iterations to remove. However, too many iterations prevents us from observing the subtle evidence in the residuals, leading to a decrease in performance.

In general, since $d=1$ provides the best results, we recommend performing only one backward diffusion step when using BTD to detect forensic anomalies.

\begin{table}[t]
\caption{Ablation Study: The impact forwards diffusion has on tamper detection. Existing methdos perform both forwards and backwards (F\&B) while BTD only performs backwards (B). The rows indicate the number of steps performed in the indicated direction(s) before measuring the residual. Bold values indicate the best result over 0.6 (close to random guessing).}
\label{tab:noise_ablation}
\begin{tabular}{cr|ccc|ccc|}
\multicolumn{2}{c|}{} & \multicolumn{3}{c|}{Injection} & \multicolumn{3}{c|}{Removal} \\
\multicolumn{2}{c|}{} & MRI & \multicolumn{2}{c|}{CT} & MRI & \multicolumn{2}{c|}{CT} \\
\multicolumn{2}{c|}{} & SD & CT-G & SD & SD & CT-G & SD \\ \hline
 & \textit{F\&B} & 0.7279 & 0.6839 & 0.7375 & 0.8408 & 0.4828 & 0.6626 \\
\multirow{-2}{*}{1} & \cellcolor[HTML]{EFEFEF}\textit{B} & \cellcolor[HTML]{EFEFEF}\textbf{0.9862} & \cellcolor[HTML]{EFEFEF}\textbf{0.8056} & \cellcolor[HTML]{EFEFEF}\textbf{0.9082} & \cellcolor[HTML]{EFEFEF}\textbf{0.9838} & \cellcolor[HTML]{EFEFEF}0.5524 & \cellcolor[HTML]{EFEFEF}\textbf{0.8607} \\ \hline
 & \textit{F\&B} & 0.5063 & 0.6502 & 0.6982 & 0.6755 & 0.4995 & 0.6086 \\
\multirow{-2}{*}{10} & \cellcolor[HTML]{EFEFEF}\textit{B} & \cellcolor[HTML]{EFEFEF}0.5873 & \cellcolor[HTML]{EFEFEF}0.6405 & \cellcolor[HTML]{EFEFEF}0.6909 & \cellcolor[HTML]{EFEFEF}0.6909 & \cellcolor[HTML]{EFEFEF}0.5061 & \cellcolor[HTML]{EFEFEF}0.6195 \\ \hline
 & \textit{F\&B} & 0.4047 & 0.6385 & 0.7062 & 0.6183 & 0.5059 & 0.6089 \\
\multirow{-2}{*}{50} & \cellcolor[HTML]{EFEFEF}\textit{B} & \cellcolor[HTML]{EFEFEF}0.4152 & \cellcolor[HTML]{EFEFEF}0.6130 & \cellcolor[HTML]{EFEFEF}0.6849 & \cellcolor[HTML]{EFEFEF}0.6058 & \cellcolor[HTML]{EFEFEF}0.5041 & \cellcolor[HTML]{EFEFEF}0.6081 \\ \hline
 & \textit{F\&B} & 0.3995 & 0.6331 & 0.7076 & 0.6025 & 0.4952 & 0.6064 \\
\multirow{-2}{*}{100} & \cellcolor[HTML]{EFEFEF}\textit{B} & \cellcolor[HTML]{EFEFEF}0.4015 & \cellcolor[HTML]{EFEFEF}0.5889 & \cellcolor[HTML]{EFEFEF}0.6895 & \cellcolor[HTML]{EFEFEF}0.5953 & \cellcolor[HTML]{EFEFEF}0.4974 & \cellcolor[HTML]{EFEFEF}0.5980 \\ \hline
 & \textit{F\&B} & 0.4266 & 0.6363 & 0.7231 & 0.6002 & 0.4930 & 0.5911 \\
\multirow{-2}{*}{200} & \cellcolor[HTML]{EFEFEF}\textit{B} & \cellcolor[HTML]{EFEFEF}0.4339 & \cellcolor[HTML]{EFEFEF}0.5685 & \cellcolor[HTML]{EFEFEF}0.6928 & \cellcolor[HTML]{EFEFEF}0.5942 & \cellcolor[HTML]{EFEFEF}0.4838 & \cellcolor[HTML]{EFEFEF}0.5754 \\ \hline
\end{tabular}%
\end{table}

\begin{figure*}[t]
    \centering
    \includegraphics[width=\textwidth]{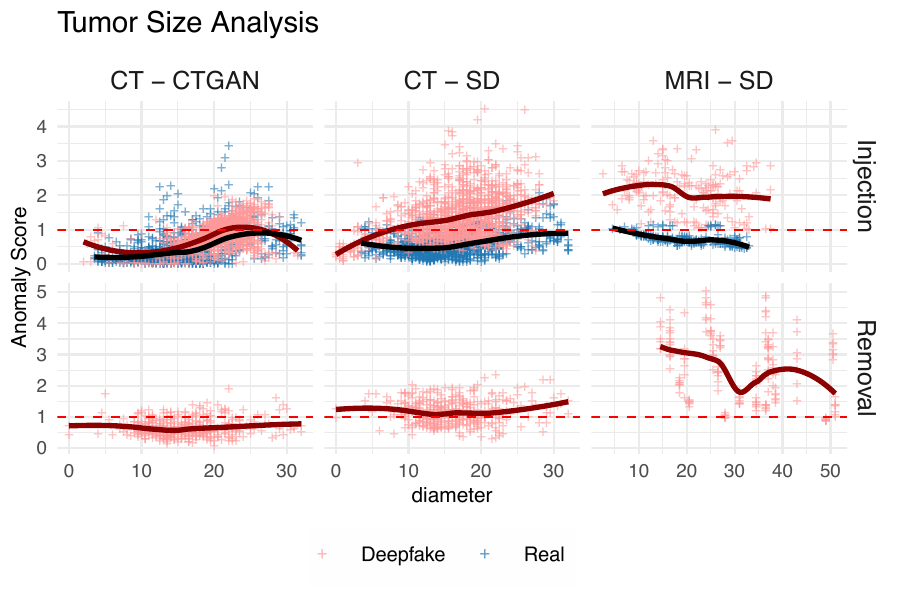}
    \vspace{-1em}
    \caption{The impact tumor size has on BTD detection performance. Anomaly scores are normalized the to a false positive rate (FPR) of 0.1 (values over the red dashed line are anomalies). In datasets of removal (bottom row), real images (TB) are omitted as they lack tumors.}
    \label{fig:tumer_size}

\end{figure*}

\subsubsection{Forensic Evidence.}
It might be assumed that larger tampered areas would be easier to detect given the presence of more evidence. While this phenomenon exists in the CT injection datasets, in Figure~\ref{fig:tumer_size} we show that the opposite is true for the MRI datasets. Upon further investigation, we found that SD tends to modify more of the image when the mask of the tumor is larger. This means that when larger masks are used, BTD has less contextual information to use to detect semantic anomalies.
This is why larger tumors are slightly harder to detect than smaller ones for MRI compared to CT where the mask size is constant. In CT removal scenarios there is no correlation to tumor  size because CTGAN always tampers the same sized area of 32x32x32, regardless of how large the tumor was. Similarly, Uniform mask sizes of 32x32 were used to remove tumors in CT with Stable Diffusion as well.

Despite these observations, we see that fake tumors have consistently higher BTD-anomaly scores, regardless of their size. 
This supports or hypothesis that BTD focuses on subtle forensic evidence and not large semantic anomalies.

\section{Conclusion}
\label{sec:conc}
In this paper, we presented Back-in-Time Diffusion (BTD), a novel approach for identifying medical deepfakes. BTD surpasses current methodologies, establishing a new benchmark for detecting tampered medical images. It exhibits robustness across a diverse set of medical imaging modalities (CT and MRI) and deepfake attacks (injection and removal).
Moreover, the method's efficacy against different deepfake generation technologies (CT-GAN and SD) further underscores its adaptability and robustness. However, despite these achievements, challenges in deepfake detection still persist. For instance, BTD, along with all other methods, struggles to detect tumors that have been removed from CT images using CT-GAN.  To assist the scientific community in addressing these challenges and to foster the development of solutions for combating medical deepfakes, we have made our datasets and source code openly accessible.\footnote{Redacted due to the double-blind review policy.} Finally, as future work, it would be interesting to explore the application of BTD to other domains, such as classical image forensics.

\begin{acks}
This material is based upon work supported by the Zuckerman STEM Leadership Program. This project has received funding from the European union's Horizon 2020 research and innovation programme under grant agreement 952172.

\begin{figure}[h]
    \centering
    \includegraphics[width=0.1\textwidth]{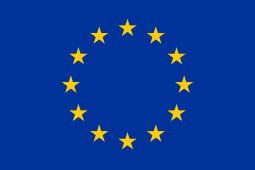}
\end{figure}
\end{acks}
\bibliographystyle{ACM-Reference-Format}
\bibliography{egbib}

\appendix





\end{document}